\documentclass{article}
\usepackage{graphicx}
\usepackage{epstopdf, epsfig}
\usepackage{amssymb}
\usepackage{latexsym}
\usepackage{amsmath}
\usepackage{mathabx}
\usepackage{amscd}

\newcommand{\nn}{\nonumber}
\def\be{\begin{equation}}
\def\ee{\end{equation}}
\newcommand{\bea}{\begin{eqnarray}}
\newcommand{\dg}{\ensuremath{^\circ}}
 \newcommand{\eea}{\end{eqnarray}}
\def\p{{\partial}}

\title{Stability of ion acoustic nonlinear waves and solitons in
magnetized plasmas}

\author{Piotr Goldstein
and Eryk Infeld
 \\{\footnotesize Theoretical Physics Division, National Centre
for Nuclear Research, Ho\.za 69, 00-681 Warsaw Poland}\\
{\footnotesize Submitted to \textit{Plasma Physics and Controlled Fusion}}}
\date{}
\numberwithin{equation}{section}

\begin{document}

\maketitle

\begin{abstract}
Early results concerning the shape and stability of ion acoustic
waves are generalized to propagation at an angle to the magnetic
field lines. Each wave has a critical angle for stability. 
Known soliton results are recovered as special cases. A historical overview of the problem concludes the paper.
\end{abstract}
\section{Introduction}
Some time ago the problem of stability of waves described by the
Zakharov Kuznetsov equation (ZK, 1974) was solved for arbitrary
shape of the wave. Its propagation, however, was limited to
clinging to the magnetic field (Infeld 1985). As a check, the
soliton case, further limited to perpendicular instabilities, was
recovered as found by Laedke and Spatschek (1982). Since then
several authors have looked at the propagation and some at
stability in various configurations (Infeld and Frycz, 1987 \&
1989, Allen and Rowlands, 1993, 1995 \& 1997, Munro \& Parks 1999, Nawaz et a1 2013, Murawski and Edwin
1992, Bas and Bulent 2010, Mothibi 2015) and several others. Here
we present a small $K$ stability analysis of a nonlinear wave
propagating at an angle to the magnetic field. Its shape is
treated exactly and a cubic is obtained for the frequency of the
perturbation (or growth rate). For zero angle, the cubic found in
Infeld 1985 is recovered, (see also the book by Infeld and
Rowlands 2000). The ZK equation is taken in the form
\be
\frac{\p n}{\p t}+\frac13\left(\p_x^2+\p_y^2+3n\right)\frac{\p
n}{\p x} = 0.
\ee
Here $x$ is along the uniform field $\boldsymbol{B}$. The wave
is propagating at an angle $\theta$ and it proves convenient to
rotate the system by this angle. Thus
\bea
&&x = \cos\theta\,\xi+\sin\theta\,\eta\nn\\
&&y = \sin\theta\,\xi-\cos\theta\,\eta,
\eea
and ZK is now
\be
\frac{\p n}{\p
t}+\left(n+\frac{\p_\xi^2+\p_\eta^2}{3}\right)\left(\cos\theta\,n_\xi
+ \sin\theta\, n_\eta\right) = 0
\ee

\section{Shape of the wave and possible perturbation}
We assume the nonlinear wave or soliton to be a
function of $\xi-Ut$

We now look for stationary nonlinear solutions of the form
\be
n_0^\prime(\xi^\prime)= n_0+U/\cos\,\theta,\quad\xi^\prime = \xi-Ut
\ee
thus adding a constant to $n_0$ which we simply include in zero order. Thus in the new variables, dropping the primes and integrating twice
\bea
&& \frac13\,\p^2_\xi n_0 +  \frac12 n_0^2 = C\nn\\
&&\frac16(\p_\xi n_0)^2  = C n_0-\frac16 n_0^3+D.
\eea
By rescaling
the variables and the constants we may always reduce the number of parameters putting
\be
C = 1/6,\quad D(C = 1/6) = \zeta/6.
\ee
To obtain positive 
\be
n_{0\xi}^2=n_0-n_0^3+\zeta
\ee
in a finite interval of $n_0$, 
the parameter $\zeta$ has to satisfy
\be
-2/\sqrt{27}<\zeta< 2/\sqrt{27}
\ee
and the stationary solution $n_0(\xi)$ is periodic with a period $\lambda$, which may be defined in terms of complete elliptic integrals.

Suppose a periodic wave solution is perturbed such that the wave
vector $\boldsymbol{K}$ of the perturbation forms an angle $\psi$
with the direction $\boldsymbol{\xi}$ of the nonlinear wave. In the coordinate
system of the basic wave we have
\bea
n&=&n_0(\xi)+\delta n = n_0(\xi)+\tilde{\delta} n(\xi) e^{i[(K\,\cos\, \psi)\xi+(K\,\sin\,\psi)\eta - \Omega t]}\\
\boldsymbol{K} &=& K\,(\cos\,\psi,~\sin\,\psi)
\eea
and $\tilde{\delta} n(\xi)$ is $\lambda$ periodic. We now assume
$K$ small and expand:
\bea
\Omega&=&\Omega_1(\psi) K+\Omega_2(\psi) K^2 +...\\
\tilde{\delta}n&=&\delta n_0+K\delta n_1+K^2 \delta n_2+...
\eea
Consistency in second order will yield a relationship of the form
\be
G(\Omega,\boldsymbol{K},\zeta)=0,
\ee
generalizing a dispersion relation theory.

Introducing
\be
L = \frac13\p_\xi^2+n_0\nn
\ee
we find
\bea
&&\p_\xi L\tilde{\delta} n=\frac13 \p_\xi^3\tilde{\delta} n+\p_\xi n_0\tilde{\delta} n =
-i \,\Omega \tilde{\delta} n \nn\\ && - 
 i K \left[\cos\,\theta\, \cos\,\psi\,( n_0\tilde{\delta} n + 
      \tilde{\delta} n_{\xi\xi} )+ 
    \sin\,\theta\, \sin\,\psi \left(n_0\tilde{\delta} n + 
       \frac13\tilde{\delta} n_{\xi\xi}\right)\right]\nn\\&&+ 
  K^2 \left[
    \cos\,\theta \left( 
        \cos^2\psi + 
        \frac13\sin^2\psi\right)+\frac23 \sin\theta\, \sin\,\psi\,\cos\,\psi 
      \right]\tilde{\delta} n_\xi
\eea
\section{Dispersion relation cubic for $\Omega/K$}
In second order in $K$, a dispersion relation is
obtained.
\bea
&&\left(\Omega ^*/K\right)^3+\frac29 \frac{[4 + 3 Y (Y + 6 \zeta)] (3 \cos\psi - \tan\theta \sin\psi)}{Y^3 - 4 Y - 8 \zeta}\left(\Omega ^*/K\right)^2\nn\\
&&+\frac{2}{15}\frac{(Y + 3 \zeta) (Y^2 - 4/3) (1 - 5 \tan^2\theta) \sin^2\psi + 
 5 [4 \zeta + Y (8/3 + 3 Y \zeta)] \sin 2 \psi \tan\theta}{Y^3 - 4 Y - 8 \zeta}\left(\Omega ^*/K\right)\nn\\
&&+\frac{8}{27}\frac{(27 \zeta^2 - 4) \cos^2\psi (\cos\psi + \sin\psi \tan\theta)}{Y^3 - 4 Y - 8 \zeta}\nn\\
&&+\frac{8}{45}\frac{(Y + 3 \zeta)^2 \sin^2\psi (\cos\psi + \sin\psi \tan\theta)}{Y^3 - 4 Y - 8 \zeta} =0
\eea
where $\Omega^* = \Omega - U K \sin\psi\tan\theta$. Here $n_1$, $n_2$, $n_3$, are
roots of $n_x^2$
in increasing order. Also $n_0$ is contained between $n_2$, and
$n_3$. Other definitions are
\be
s = \frac{n_3-n_2}{n_3-n_1}\le 1,\quad Y=2\langle n_0\rangle =
2\left[n_1+(n_3-n_1)\frac{E(s)}{K(s)}\right]
\ee
$E(s)$ and $K(s)$ are complete elliptic integrals of modulus $s$.

For $\theta=0$ we regain the result of the Infeld and Rowlands book (eq. (8.3.90) in ref. [8]).

We limit further analysis to the most unstable angle of perturbation $\psi=\pi/2$.

There is an instability for $\theta=0$, and $\theta>0$ up to a critical angle. The angle may easily be obtained e.g. by an analysis of the discriminant of the cubic. For $\zeta=0$ ($n_1=-1,~n~_2=0,~n_3=1$), the critical angle is equal 
$0.792946~\mathrm{rad}\approx 
45.4325$\dg~(Fig.1).
\begin{figure}[htbp]\label{Figure1}
\begin{center}
\includegraphics[totalheight=12cm]{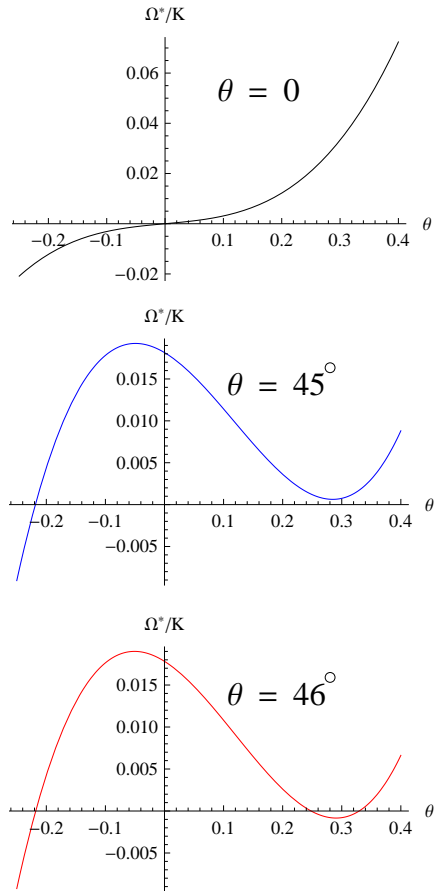}
\caption{The cubic for $\theta=0,~\theta=45$\dg~and $\theta=46$\dg, while $\psi=\pi/2$ . Single $\theta$-intercept means that two roots are complex conjugate; one of them corresponds to the unstable mode. When the cubic has 3 real roots, no instability occurs. The critical angle apparently lies between 45\dg~and 46\dg.}
\end{center}
\end{figure}

The growth rate $\Gamma/K$ increases from its value at $\theta=0$ to a maximum, then falls to zero at the critical $\theta$ (Fig.2). For all acute angles $\theta$ above the critical one the system is stable to first order in $K$. 

For $\theta>\pi/2$ the situation is symmetric with respect to 
$\theta\rightleftarrows \pi-\theta$.
\begin{figure}[htbp]\label{Figure2}
\begin{center}
\includegraphics[totalheight=6cm]{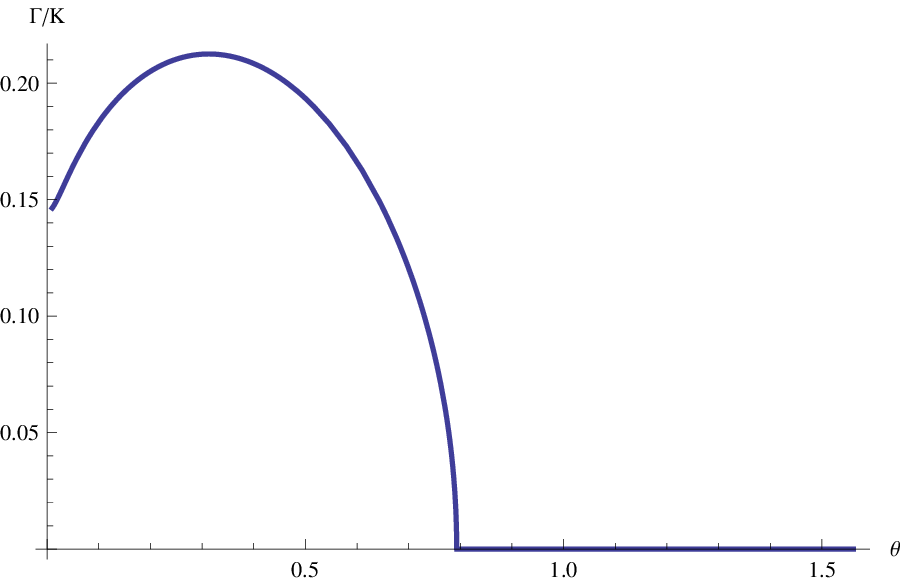}
\caption{The growth-rate $\Gamma/K$ as a function of $\theta$ for $\zeta=0$. The instability vanishes above the critical angle $\theta\approx 0.793$.}
\end{center}
\end{figure}

\begin{figure}[htbp]\label{Figure3}
\begin{center}
\includegraphics[totalheight=6cm]{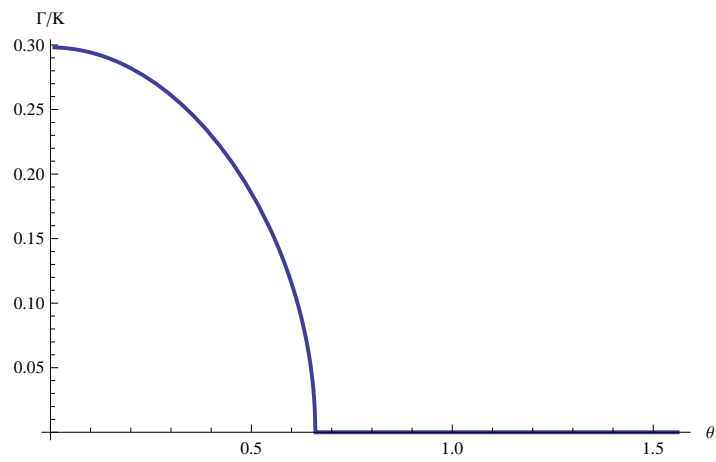}
\caption{The growth-rate $\Gamma/K$ as a function of $\theta$ for the soliton case $\zeta=2/\sqrt{27}$. The growth-rate decreases from 0.298 at $\theta=0$ to zero at the critical angle  $\theta\approx 0.659$.}
\end{center}
\end{figure}

Another limit of interest is for a soliton propagating at an angle 
to $\boldsymbol{B}$. For it $\zeta= 2/\sqrt{27}$
and the cubic dispersion relation reads
\be
(\Omega^*/K)^3 - 
  \frac{\sin\theta}{3 \sqrt{3}} (\Omega^*/K)^2 + 
 \frac{4}{45}(\cos^2\theta- 5 \sin^2\theta) (\Omega^*/K)
  +  \frac{4}{45 \sqrt{3}} \cos^2\theta\sin\theta=0
\ee
Here the growth rate is greatest for $\theta=0$, where it is equal $2/(3 \sqrt{5})\approx 0.298142$. Then it decreases to zero at the critical angle, which is
 $\mathrm{arctan}\sqrt{3/5}
 \approx 0.659058~\mathrm{rad}\approx 37.7612$\dg (Fig.3), in agreement with Allen and Rowlands (1995).
 To first order in $K$ the system is stable for acute angles greater than this angle. The behaviour for angles above $\pi/2$ again follows from the symmetry $\theta\rightleftarrows \pi-\theta$.

\section{A bit of history}
Forty years ago Infeld and Rowlands pointed out flaws in the way people were
calculating the stability of solitons. The perturbations introduced failed to vanish
at infinity (1977). Many scientists took the problem seriously. The two authors
looked at the problem differently. Rowlands pointed out that the soliton problem
involves four different kinds of secularity and removed all by introducing multi
variables. Infeld pointed out that removing secular terms is simpler for periodic structures,
so lets treat a soliton  train and take it to the $\lambda\to\infty$ limit. (This effort is in
that spirit.) Lowest order results of the two methods so far agree, but the transition
is not understood. All this notwithstanding the nonlinear wave problems' importance.

\end{document}